\documentclass{article}
 \usepackage{float}
\usepackage{graphicx}
\usepackage{amsmath, amsthm, amsfonts,ams symb}
\usepackage[symbol]{footmisc}

\theoremstyle{definition}

\theoremstyle{remark}

\usepackage{natbib} 
\title{Three arguable concepts: 
  point particle singularity,   
  asymmetric action of EM on quantum wave functions, and 
  the Left out restricted Lorentz gauge from U(1)
  \footnote{Dedicated to the International Year of Basic Sciences for Sustainable Development (IYBSSD) 2022.}}
\author{Yousef Sobouti\\
  \small Institute for Advanced Studies in Basic Sciences\\
   sobouti@iasbs.ac.ir\\https://orcid.org/0000-0002-4356-1460\\
  \small 444 Prof. Yousef Sobouti Blvd.,\\
  \small Zanjan 45137-66731, Iran\\
 }

\begin{document}
\date{}
\makeatother
\maketitle
\abstract{We address three concepts. 
i. The point particle assumption inherent to non-quantum physics is singular and entails divergent fields and integrals. 
ii. In quantum physics electromagnetism (EM) plays an asymmetric role. It acts on quantum wave fields   (wave functions) but the  wave fields do not react back.
We suggest to promote the one sided action of EM  on quantum waves into a mutual action-reaction  partnership.  By so doing, the quantum wave shares its analyticity   with the  EM field and removes the latter's singularities and divergences.   
iii. The conventional U(1) symmetry  leaves quantum dynamics invariant under a 'general' Lorentz gauge and imposes the standard minimal coupling of the quantum wave to the EM 4-vector potential.  One, however, has the option to ask for invariance under the 'restricted' Lorentz gauge. This in turn invites in a coupling to the derivatives of the vector potential in addition to the minimal coupling and, so to say, enlarges the U(1) symmetry.   
We examine the Dirac electron in this context and find that the electron exhibits distributed charge and current densities. The enlarged symmetry is expected to bring in its own constant of motion.   Indeed it does. The anomalous g-factor of the so designed electron emerges, up to order $(\frac{\alpha}{\pi})^2$ as the new constant of motion in agreement with the QED theorized values.\\

}

 \section{Introduction}
 
 i)  A tacit assumption in  non-quantum physics, be it Newtonian dynamics, electromagnetism, relativities, etc., is the concept of point particles, infinitely small volumes packed with finite amounts of charges, masses, energies, etc. It is true that, as a working hypothesis, the concept is capable of  coping  with most of the practical eventualities. But the concept is singular, defies the common sense, and does not conform to the practised code of physics, that any thing talked about should, in principle, be measurably verifiable.
 
 Since the inception of Newton's laws of motion and gravitation or of Coulomb's law of the electric field,  inquisitive minds time and again have raised the issue of singularities  and have offered solutions.  Examples from the earlier days  are \citeyear{born1934foundations}, \citeyear{lande1941finite},  \citeyear{podolsky1948228}, \citeyear{green1947self}. 
    In the very recent years, and long after the emergence of the QED, \cite{blinder2003singularity} suggests constitutive properties to vacuum in the ultra-microscopic vicinity of charged particles. \cite{santos2011plasma} expands on Podolsky's regularized electrodynamics and plasma-like vacuum.   \cite{lazar2020second} propose a second gradient electromagnetostatics. 
  
ii) Asymmetric action of EM on Quantum wave  fields: 
Quantum mechanics deals with waves and wave packets and offers but partial solutions. Partial in the sense that, for instance, in the hydrogen atom the motion of the electron is wave-like, but its own charge and that of the proton sitting at the center is point-like.  
There is an asymmetry here.  The EM field  acts  on the quantum wave  field, but the wave field does not react back. Can one follow \cite{einstein2014meaning} and say that
 \textit{it is contrary to the mode of thinking in science to conceive of a thing that acts itself, but which cannot be acted upon.}
 
 In this paper we consider a charged particle. On account of its being a quantum entity, it possesses a wave function, a quantum wave field (QW).   And on account of its being a charged one, it has an EM field.  
  We conjecture, a la Einstein, a mutual interaction between the EM and the QW field, by simply writing down their combined Lagrangian densities  and deriving the relevant Euler- Lagrange equations.    
Quantum fields, on account of the uncertainty principles, are singularity free. Upon acting  on other fields share this feature with them, here the  EM field of the particle. We show that the singular Coulomb potential becomes  quadratic at the origin, as if the charge of the electron is spread out and is  no longer point like. Considering electron's spin, however, a rotating  spread- out charge is an electric current system and thereof the electron acquires a self induced magnetic moment.     The electric  field at far distances, however,  remains Coulombian,  but gets  screened by a small factor.  

iii) Absence of the restricted Lorentz gauge from  U(1):
 The Pauli-Schroedinger or the  Dirac wave fields are U(1) symmetric. That is, if the wave field is multiplied by a spacetime dependent phase factor and EM is written in  Lorentz gauge the EM and QW equations remain invariant. This requires the minimal coupling of the QW field to the EM vector potential,  $A_\mu$. One, however, has still the option to ask for symmetry under  the restricted Lorentz gauge. This in turn invites in a coupling to derivatives of the 4-potential, $\partial_\nu A_\mu$.
  
 As a pedagogical example, in the introduction of section 2 below, we briefly examine the case of a non-relativistic spineless charged particle to demonstrate how the mutual EM-QW coupling is capable of removing  Coulomb- like singularities.  The bulk of the section 2, however, is devoted to a detailed analysis of the Dirac electron in the context  elaborated  above. We come up with a set of coupled Euler- Lagrange equations for the EM and the QW fields.  In section 3 we attempt an iteration solution of these equations. The highlight of the paper is the subsection 3.3, where we analyse the anomalous g- factor of our proposed electron. Up to the order $(\frac{\alpha}{\pi})^2$ we  find it in concordance with  the QED- theorized and the laboratory measured values. We present this concordance as an observable  evidence to support our conjectures.    
 Naturally, any  new assumption raises new questions. A short list of such questions is narrated in   section on Concluding Remarks.  

\section{A Lagrangian formulation} 
 As an introduction let us consider a charged particle with a complex Schroedinger-like QW field $\psi(t, \mathbf{x})$ and a static electric potential field $A_0(\mathbf{x})$. For the Lagrangian density of both fields and their mutual interaction let us  write 
 \begin{eqnarray}
{\cal L} = \psi^*\left[-i\hbar \frac{\partial}{\partial t} 
- \frac{\hbar^2}{2m}\nabla^2  + zeA_0 \right]\psi  - \frac{1}{2} z|\nabla A_0|^2.   \label{LpsiV}
\end{eqnarray} 
The first two terms in (\ref{LpsiV}) are the Lagrangian density of the free field $\psi$ The last term is that of the free field $A_0$.  The third term, $ze\psi^*A_0\psi$, is the  interaction Lagrangian, where  $z$ is a dimensionless constant that couples  the two fields together. 
The Euler-Lagrange equations  for $\psi$ and $A_0$ are as follows,
\begin{eqnarray}  
&& - i\hbar\frac{\partial\psi}{\partial t} - \frac{\hbar^2}{2m}\nabla^2\psi + zeA_0\psi = 0, 
\label{psiShrod}\\
&& ~~ \nabla^2 A_0 = -e\psi^*\psi. \label{VPoisson}
\end{eqnarray}
Equation(\ref{psiShrod}) is a Schroedinger equation for $\psi$ in an EM field given by (\ref{VPoisson}). The latter is a Poisson equation for $A_0$, where  $e\psi^*\psi$ serves as the charge density. It is seen that the charge density is not a  delta function distribution. Thus, $A_0(\mathbf{x})$ will not be Coulomb-singular. The total charge  is $-e$, for the $\psi$ is normalized to $1$. It is easy to show that the potential $A_0(\mathbf{x})$ at distances of the order of Bohr radius is quadratic at some constant potential well, while it remains Cloulombian  at far distances. The two  EM and QW fields are coupled together. The linearity of both equations is lost. One may, however, proceed with an iteration scheme and produce approximate solutions. More on such issues will be addressed in the coming  section. 
 
 \subsection{Dirac Electron}

Our notation in this section is that of \cite{sakurai2006advanced}.
The EM unit system is the Heaviside - Lorentz one. 

 We consider a spin $1/2$  electron and conjecture a mutual interaction between its Dirac QW  and EM fields.   We  find  a) the  electron acquires a charge distribution and  the electric field becomes non-singular. b) A spinning distributed charge is a current system. Therefore, the electron develops  a finite self induced magnetic moment with an anomalous g-factor correct to order $(\alpha/\pi)^2$.  Both features  are reminiscent of what one finds in the renormalized and scale invariant systems, but without resorting to the QED formalism.\\

 \noindent The Lagrangian density to deal with is.
\begin{eqnarray}
 {\cal L} &=& 
  \bar{\psi}\left[\gamma_\mu \left( \partial_\mu - \frac{ize}{c\hbar}A_\mu \right) 
  + \frac{ize}{4c\hbar}~(\kappa\alpha^2 a_0) ~\gamma_\mu\gamma_\nu \partial_\nu A_\mu + \frac{mc}{\hbar} \right] \psi \nonumber\\
  &&  +  \frac{z}{4 c\hbar} F_{\mu\nu}F_{\mu\nu}, \label{LDirac}
 \end{eqnarray}
  where  
   \begin{eqnarray} 
  &&  \gamma_\mu\textrm{'s are Dirac matrices},\nonumber\\
&& \bar{\psi} = \psi^\dagger\gamma_4, \nonumber\\
&& x_\mu = (\textbf{x}, x_4=ict), \nonumber\\
&& A_\mu = (\textbf{A}, A_4= iA_0),~\textbf{A}~\textrm{vector potential}, ~~ 
  A_0 ~~\textrm{scalar potential},\nonumber\\
&& F_{\mu\nu} = \partial_\mu A_\nu - \partial_\nu A_\mu. \nonumber\\
&& \nonumber\\
&& a_0 = \frac{4\pi\hbar^2}{m e^2} = \frac{\bar{\lambda}_C }{\alpha} 
                          = 0.529 \times 10^{-10} m, \textrm{Bohr radius}, \nonumber\\
&& \bar{\lambda}_C = \frac{\hbar}{mc} 
                          = 0.243\times 10^{-11}m, \textrm{reduced Compton wavelength},\nonumber\\
 && \alpha =  \frac{ e^2}{4\pi\hbar c} \approx \frac{1}{137}, \textrm{fine structure constant}.\nonumber
\end{eqnarray}
The first and fourth terms in (\ref{LDirac}) are the Lagrangian density of the free Dirac field.  The last term is that of the free (i.e. the source-less) EM field.  The second term is the conventional minimal interaction Lagrangian of the Dirac wave with the 4-vector potential $A_\mu$. 
The third term is yet another interaction involving the derivatives of the vector potential, $\partial_\nu A_\mu$.  We shall  come back to it shortly below.  
The $z$ and $\kappa$ are two as yet unspecified dimensionless coupling constants.  They will be decided later by requiring  our calculated g- factor to be con-formant with the laboratory measured  and/or the QED-derived counterparts.  
The    factor  $\alpha^2 a_0 = \alpha \bar{\lambda}_C$ in the fourth term and $1/c\hbar$ in the other terms are so chosen to make  all terms in $\cal L$ to have the same physical dimension. \\

\subsection{Symmetries of the Lagrangian (\ref{LDirac})}

Dirac electron minimally coupled to  an EM field is said to be $U(1)$ symmetric, meaning that the  Lagrangian of (\ref{LDirac}) (without the $\kappa$ term) remains invariant under the transformation,
$$ \psi' = \exp(ie\chi/c) \psi, ~~ A'_\mu = A_\mu +\partial_\mu \chi, ~~ \chi(x) ~~\textrm{arbitrary.}$$
  An arbitrary Lorentz gauge, however, does not exhaust all redundancies in the choice of $A_\mu$.  One still has the option to choose a restricted  gauge $\chi$ with $\square^2 \chi=0$ and come up with the same Dirac-  and the same EM- fields. But insisting on invariance under the restricted gauge invites in a coupling of  $\psi$ to derivatives of the vector potential, $\partial_\nu A_\mu$, namely the $\kappa$ term of (\ref{LDirac})  in addition to the minimal coupling.
To see that the $\kappa$ term is invariant under the restricted  gauge, let 
$$ \psi' = \exp(ie\chi/c) \psi, ~~ A'_\mu =A_\mu + \partial_\mu \chi, ~~ \square^2 \chi = 0,$$
and examine the  following, 
\begin{eqnarray}
&& (\bar{\psi'} \gamma_\mu\gamma_\nu \psi') \partial_\nu A'_\mu 
= \exp(-ie\chi/c)(\bar{\psi}\gamma_\mu\gamma_\nu\psi) \exp(ie\chi/c)\partial_\nu(A_\mu + \partial_\mu \chi) \nonumber\\
&&\hspace*{2.4cm} =  (\bar{\psi} \gamma_\mu\gamma_\nu\psi) \partial_\nu A_\mu 
+ ( \bar{\psi} \gamma_\mu\gamma_\nu \psi) \partial_\nu \partial_\mu\chi \nonumber\\
&& \hspace*{2.4cm} =  (\bar{\psi} \gamma_\mu\gamma_\nu\psi) \partial_\nu A_\mu 
+\bar{\psi} \psi \square^2 \chi \nonumber\\
&& \hspace*{2.4cm} =  (\bar{\psi} \gamma_\mu\gamma_\nu\psi) \partial_\nu A_\mu.  
\label{Rest.gauge}
\end{eqnarray}
In (\ref{Rest.gauge}), on the right hand side of the second equality, the factor 
$(\bar{\psi} \gamma_\mu\gamma_\nu \psi)$ is antisymmetric and the factor $\partial_\nu \partial_\mu\chi$ is symmetric on swapping $\mu\leftrightarrow \nu$. Their product vanishes unless $\mu=\nu$, which then reduces to $\square^2 \chi=0$.  Thus,   the $\kappa$ term in (\ref{LDirac}) is an invariant of the motion.  We will come back below and show that the Noether constant associated with it is the $\alpha/2\pi$ anomalous  g- factor of the self induced magnetic moment of the electron.\\
 
 \begin{itemize} \begin{small}
\item[]
Digression: The $\kappa$ interaction in (\ref{LDirac}), in the form of  $\bar{\psi}\gamma_\mu\gamma_\nu\psi   F_{\mu\nu}$,
 has precedence in the pre- QED literature and is known as the anomalous Pauli interaction. 
As a phenomenological term,  it was introduced  to  give the $\frac{1}{2}(\frac{\alpha}{\pi})$ anomalous magnetic moment of the electron. It went, however,  into oblivion once Schwinger obtained the same result in 1948, via QED.
 Here we argue that  the $\kappa$ interaction  is a genuine  symmetry based interaction.  And as we will see below, its predictions of the anomalous magnetic moment go beyond   Schwinger's correction. We propose it to be considered  as an alternative to the QED way of looking into the problem. \end{small}  \end{itemize}   
 
\subsection{Reduction of the Euler- Lagrange equations}
 
Coming back to (\ref{LDirac}), the two  equations for $\psi$ and $F_{\mu\nu}$ are,
\begin{eqnarray}
&&\gamma_\mu\left(p_\mu - \frac{ze}{c}A_\mu\right)\psi - i mc \psi  
             + \frac{1}{4}\kappa z \alpha^2 a_0 e\frac{1}{c}
                      \frac{\partial A_\mu}{\partial x_\nu }\gamma_\mu \gamma_\nu \psi = 0,  
 \label{psiDirac}\\ 
 \nonumber\\
 &&-\frac{\partial F_{\mu\nu}}{\partial x_\nu} = \square^2 A_\mu
    = -  i e\bar{\psi}\gamma_\mu\psi 
  - i \frac{1}{4}\kappa\alpha^2 a_0 e\frac{\partial}{\partial x_\nu}\left(\bar{\psi} \gamma_\mu\gamma_\nu \psi\right).
   \label{FmunuDiracCoupled}
\end{eqnarray}
  Noting that
\begin{eqnarray}
\gamma_4 = \left[\begin{array}{c c} 
1 & 0              \\ 0           & -1\end{array}\right],~~  
\gamma_i= \left[\begin{array}{c c} 
0 & -i\sigma_i \\ i\sigma_i &  0\end{array}\right] ~~~~\textrm{and}~~
\gamma_4\gamma_i = -i\left[\begin{array}{c c} 
0 & \sigma_i     \\ \sigma_i      & 0\end{array}\right],
\nonumber
\end{eqnarray}
one may split (\ref{FmunuDiracCoupled}) for $A_\mu$ into its scalar and vector components, 
\begin{eqnarray}
 \square^2 A_0   
    &=& - e \psi^\dagger\psi  \nonumber\\
    &&  + i\frac{1}{4}\kappa \alpha^2 a_0 e\frac{1}{c} \frac{\partial}{\partial t} \left(\psi^\dagger\gamma_4\psi\right)
    \nonumber\\
          && - i \frac{1}{4}\kappa\alpha^2 a_0 e\frac{\partial}{\partial x_i} \left(\psi^\dagger \left[\begin{array}{c c} 0 & -\sigma_i \\ \sigma_i & 0 \end{array}\right]\psi\right), 
          \label{A0scalar}\\
\nonumber\\
 \square^2 A_i   &=& - e\psi^\dagger 
 \left[\begin{array}{c c} 0 &\sigma_i \\ \sigma_i & 0 \end{array}\right] \psi 
 \nonumber\\
&& +i \frac{1}{4}\kappa \alpha^2 a_0 e\frac{1}{c}\frac{\partial}{\partial t} 
\left(\psi^\dagger \left[\begin{array}{c c} 0 &-\sigma_i \\ \sigma_i & 0 \end{array}\right]\psi\right) \nonumber\\
&& +\frac{1}{4}\kappa \alpha^2 a_0 e ~\epsilon_{ijk} \frac{\partial}{\partial x_j} \left(\psi^\dagger\left[\begin{array}{c c} \sigma_k & 0 \\ 0 & -\sigma_k\end{array}\right]\psi\right).
\nonumber\\
 \label{Aivector}
\end{eqnarray}
Equations (\ref{psiDirac}) - (\ref{Aivector}) are mutually coupled and are non-linear. In an iteration scheme, one may assume a reasonable expression for the EM field, substitute it in (\ref{psiDirac}) and  solve it for $\psi$, substitute the result back in (\ref{A0scalar}) and (\ref{Aivector})  and solve them for improved  $A_0$ and $\mathbf{A}$,  go back and repeat the cycle.  
 
 \section{Iteration carried out}
To begin with let us ignore the $\kappa$ term in (\ref{psiDirac}) and for the starting EM field assume that of the classical electron, 
\begin{eqnarray}
A_0 = - \frac{e}{4\pi r}, ~~~\textbf{A} = 0.  \nonumber
\end{eqnarray}
Equation (\ref{psiDirac}) reduces to the classic equation  for a hydrogen-like atom with a relativistic spin $1/2$ electron,
\begin{eqnarray}
i\hbar\frac{\partial \psi}{\partial t} &=&\left[  c\hbar \gamma_4\gamma_i \frac{\partial}{\partial x_i} 
                                                         + \gamma_4 m c^2 - z\frac{e^2}{4\pi r}\right]\psi. \label{HDirac} 
    \end{eqnarray}
Equation (\ref{HDirac}) has been known since late 1920s. Its solutions can be found in the classical quantum mechanical and spectroscopic books, including \citep{schiff1955quantum} and\citep{sakurai2006advanced}. Its ground state solution, copied from Sakurai is,
\begin{eqnarray}
\psi &=& \left[\frac{N^2}{\pi(a_0 /z)^3}\right]^{1/2}\exp\left(\frac{-zr}{ a_0}\right) \left(\frac{2z r}{ a_0}\right)^{-\beta^2/2}  
                   \left[\begin{array}{c} 
                                             \chi \\
                                         \frac{i\beta^2}{2z\alpha}\frac{x_j}{r}  \sigma_j \chi 
      \end{array}\right],  \label{psigdDirac}
  \end{eqnarray} 
 where $\alpha$ and $a_0$ are  as before  the fine structure constant and the Bohr radius, respectively,                                               \begin{eqnarray}
 \beta^2 &=& 2\left(1 -\sqrt{1-(z\alpha)^2}\right) \label{betasquare}\\
 &=&  (z\alpha)^2 \left(1+\frac{1}{4}(z\alpha)^2 + \frac{1}{8}(z\alpha)^4 \right)
 +{\cal O}(z\alpha)^8,\nonumber\\
 N^2 &=&  \frac{2\left(1-\frac{1}{4}\beta^2\right)}{\Gamma(3-\beta^2)}. \label{Nsquare}
   \end{eqnarray}
  and $\chi$ is a Pauli  spinor, either 
                                            $$\chi^+ =  \left[\begin{array}{c}
                                               1 \\
                                               0\end{array}\right], 
                                               ~~~\textrm{or}~~~
                                               \chi^- = \left[\begin{array}{c}
                                               0 \\
                                               1\end{array}\right],$$   
  satisfying
  $$ \sigma_3\chi^\pm = \pm \chi^\pm, ~~~\sigma_3 ~\textrm{the Pauli matrix.} $$\\
  \noindent
The normalization constant  $N$  is chosen to have 
$$\int \psi^\dagger\psi d^3x=1.$$
  The combinations $z\alpha$, $a_0/z$ and $2zr/a_0$   appear frequently in the calculations. To economize in writing we introduce the following  shorthand notations,
\begin{eqnarray}
\bar{\alpha} = z\alpha, ~~~~ \bar{a}_0 = \frac{a_0}{z}, ~~~~ u=\frac{2zr}{a_0}, ~~~~
\frac{\partial}{\partial x_j} = \frac{2}{\bar{a}_0}\frac{u_j}{u}\frac{\partial}{\partial u}, \nonumber
\end{eqnarray} 
and rewrite (\ref{psigdDirac}) anew, 
\begin{eqnarray}
\psi &=&  \left(\frac{N^2}{\pi\bar{a}_0^3}\right)^{1/2}e^{- u/2} u^{-  \beta^2/2}  
                   \left[\begin{array}{c} 
                                             \chi \\
                                         \frac{i\beta^2}{2\bar{\alpha}} \frac{u_j}{u}  \sigma_j \chi
      \end{array}\right].  \label{psigdDiracbis}
  \end{eqnarray}

 \subsection{Reduction of the scalar potential $A_0$, (\ref{A0scalar})}
 The  $\psi$ of (\ref{psigdDiracbis}) is an eigenfunction with an exponential time dependence. Therefore, the EM fields $A_0$ and $A_i$ become time independent and $\square^2\rightarrow\nabla^2$. Both $\kappa$ terms in  (\ref{A0scalar}) vanish, the first one because of its time independence, and the second   because of the off diagonal $\sigma$'s in its argument.  
 Substitution of (\ref{psigdDiracbis}) in (\ref{A0scalar}) gives,
\begin{eqnarray}
 \frac{\bar{a}_0^2}{4}\nabla^2 A_0  &=&
   \nabla_u^2 A_0  = -\frac{\bar{a}_0^2}{4} e\psi^\dagger\psi  
     =   - \frac{N^2 e}{4\pi \bar{a}_0}    e^{-u} u^{-\beta^2}.  
       \label{NablaA0Dirac}
\end{eqnarray} 
   The right hand side of (\ref{NablaA0Dirac}) is the charge density responsible for the generation of $A_0$. Its integral, the total charge, is the bare charge of the electron, $-e$.  The $\delta$-function distribution of the Coulomb case, is replaced by the mild logarithmically divergent    $u^{-\beta^2} = -\beta^2 \ln u$, as $u\rightarrow 0$,  reminiscent of the rescaling practice and the logarithmic divergences of QED.  The integral of (\ref{NablaA0Dirac}) is, 
   \begin{eqnarray}
    A_0 &=&  - \frac{N^2 e}{4\pi \bar{a}_0}              
        \left[\frac{1}{u}     \Gamma(3 - \beta^2,u) - \Gamma(2-\beta^2,u) +\Gamma(2-\beta^2)\right], 
              \label{A0Dirac}
  \end{eqnarray}
    where  $\Gamma(s, u)$'s and $\Gamma(s)$  are the incomplete and complete $\Gamma$ functions, respectivly.
\begin{eqnarray}    
\Gamma(s+1,u) &=& \int_0^u \exp(-u) u^s du = s\Gamma(s,u) - \exp(-u) u^s,  \nonumber\\
\Gamma(s+1) &=& \int_0^\infty \exp(-u) u^s du = s\Gamma(s).  \nonumber
\end{eqnarray} 
In the Appendix B  we have elaborated on $\Gamma$ functions for integer and non integer numbers $s$ and have given their limiting values for small and large $u$'s.  
  The far distance limit of $A_0$ is easy to get. As $u\rightarrow \infty$, the incomplete $\Gamma$'s become complete and (\ref{A0Dirac}) reduces to 
  \begin{eqnarray}
  A_0 \rightarrow -\left(1-\frac{1}{4}\beta^2\right)\frac{e}{4\pi r},~~\textrm{as} ~~r\rightarrow \infty. \label{A0DiracInfty} 
  \end{eqnarray}
  The far distance electric potential is Coulombian. Although the total charge is $-e$, the effective potential is screened by the factor 
  $(1-\frac{1}{4}\beta^2)\approx (1- \frac{1}{4} \bar{\alpha}^2)$.  
  This is a new feature.  We will come back to it  in section on  Concluding Remarks and suggest a parallelism between this factor that aberrates the  electric field and the anomalous g- factor that plays similar role in determining the magnetic moment of the electron.\\

To get the near distance behaviour of $A_0$, one has to expand $\Gamma$'s for $u\rightarrow 0$.  With the help of (\ref{3betau}) and  (\ref{2betau})  of the Appendix B we find
 \begin{eqnarray}
 A_0 &=& -\frac{e}{4\pi \bar{a}_0}
  \left[\left(1 +\frac{1}{4}\beta^2\right) 
 - \frac{1}{6}\left(1- \frac{3}{4}\beta^2\right) u^2 +\frac{1}{6}\beta^2 u^2  \ln u\right]. \label{A0Dirac,limit0}
 \end{eqnarray}
 The first term is a potential well of finite depth. The second one  is a quadratic potential. The third term is a small (note $\beta^2\approx \bar{\alpha}^2$) quadratic-logarithmic one. We note that, in spite of the mild logarithmic singularity in the charge distribution, $A_0$ is non- singular everywhere.    
 
\subsection{Reduction of the vector potential $A_i$, (\ref{Aivector})} 
 
\textbf{Preliminaries:} 
 We recall
$$ u=\frac{2}{\bar{a}_0} r, ~~~~ \nabla^2 = \frac{4}{\bar{a}_0^2}\nabla_u^2, ~~~~
 \frac{\partial}{\partial x_j} =  \frac{2}{\bar{a}_0}\frac{u_j}{u}\frac{\partial}{\partial u},$$
 where $u_j$ is the $j$th component of $\mathbf{u}=2\mathbf{r}/a_0$.   
In the spherical polar coordinates one has,
$$\frac{u_1}{u}= \sin\theta\cos\phi 
= -\frac{1}{2}\left(\frac{8\pi}{3}\right)^{1/2}\left[Y_1^1(\theta,\phi)-Y_1^{-1}(\theta,\phi)\right],$$
$$\frac{u_2}{u}= \sin\theta\sin\phi 
= -\frac{1}{2i} \left(\frac{8\pi}{3}\right)^{1/2}\left[Y_1^1(\theta,\phi)+Y_1^{-1}(\theta,\phi)\right].$$
Note also the Cartesian components of the unit vector $\hat{\phi}$ at the point $(\theta, \phi)$, 
$$\hat{\phi}  = - \hat{x}~\sin\phi +\hat{y}~\cos\phi.$$
Returning to (\ref{Aivector}),  the second term vanishes on account of its argument being time-independent.   In the surviving terms, substituting for $\psi$ from (\ref{psigdDirac}), one gets,
\begin{eqnarray}
\nabla^2 A_i  
          &=&  \frac{N^2\beta^2}{ \pi \bar{a}_0^3 \bar{\alpha}} e \left[e^{-u}u^{-\beta^2}     
            -\frac{1}{2}\kappa  \bar{\alpha}\frac{\partial}{\partial u} \left(e^{-u} u^{- \beta^2} \right) \right] \epsilon_{ij3}\frac{u_j}{u}.      
\label{nablaAiDirac} 
 \end{eqnarray}
  The term $\epsilon_{ij3}$ pops up in the course of reducing the following expression, 
 $$\chi^\dagger(\sigma_i\sigma_j)\chi  
        = i\epsilon_{ijk} \chi^\dagger\sigma_k\chi
        = i\epsilon_{ij3} \chi^\dagger\sigma_3\chi = i \epsilon_{ij3}.$$
  The last two equalities follow from the fact that   $\chi$'s are chosen as the eigenspinors of $\sigma_3$. 
   Noting that   $u_j$ is the $j$th Cartesian component of the vector  $\mathbf{u}$, one finds
 $$ \epsilon_{123}\frac{u_2}{u} = \sin \theta \sin\phi ~~~~~\textrm{and }~~~~~ 
     \epsilon_{213}\frac{u_1}{u} = -\sin \theta \cos\phi.$$
The vector potential has no $z$- component and in the $(x,y)$- plane  lies in the $\phi$- direction. 
   Thus,
   \begin{eqnarray}
 && \nabla^2 \mathbf{A}  = \frac{1}{c}\mathbf{J}
       =   \frac{N^2\beta^2}{ \pi \bar{a}_0^3\bar{\alpha}} e    \label{nablaAphi}  \\
 && \hspace{0.8cm}    \times \left[\left(1+ \frac{1}{2}\kappa\bar{\alpha}\right)e^{-u}u^{-\beta^2} 
                   +\frac{1}{2}\kappa\bar{\alpha}\beta^2 e^{-u}u^{-(1+\beta^2)}   
                          \right] \sin\theta \hat{\phi}. 
  \nonumber                         
 \end{eqnarray} 
  The 3-vector $\mathbf{J}$ introduced above is to be considered as the  current density responsible for the creation of the vector potential. It satisfies the continuity equation,  
   $$\nabla. \mathbf{J =0}.$$
   The associated  magnetic moment density is
   \begin{eqnarray}
&&{\cal \mathbf{M}}(u)=\frac{1}{2 c} \mathbf{r}\times  \mathbf{J}(u) 
=\frac{\bar{a}_0}{4 c} \mathbf{u}\times  \mathbf{J}(u)  \label{MagMomDensity} \\
 && \hspace{.3cm}= \frac{N^2\beta^2}{4\pi \bar{a}_0^2\bar{\alpha}} e   
                         \left[\left(1+ \frac{1}{2}\kappa\bar{\alpha}\right)e^{-u}u^{-\beta^2} 
                   +\frac{1}{2}\kappa\bar{\alpha}\beta^2 e^{-u}u^{-(1+\beta^2)}\right]   
                      \sin\theta ~\mathbf{u} \times \hat{\phi}. \nonumber    
\end{eqnarray}   
As in the charge density of (\ref{NablaA0Dirac}), note the weak logarithmic divergence of ${\cal \mathbf{M}}(u)$ hidden in $u^{-\beta^2}$ in the second term of (\ref{MagMomDensity}). As for its vectorial character, expressed in Cartesian coordinates, one has, 
$$\mathbf{u}\times \hat{\phi}= u (-\sin\theta\cos\theta\cos \phi~\hat{x} 
+\sin\theta\cos\theta\cos \phi~\hat{y} +\sin^2\theta ~\hat{z}).$$
The space integral of (\ref{MagMomDensity}) is the total self induced magnetic moment.  Its $x$ and $y$ components vanish on account of their $\phi$- dependence.  The  $z$ component gives,    
\begin{eqnarray}
\mu_e \hat{z} &=&  \int \mathbf{M}d^3x \nonumber\\
&=& \frac{1}{2}\bar{a}_0 \bar{\alpha} e 
 \left[ 1+ \frac{1}{2} \kappa \bar{\alpha} - \frac{1}{3} \beta^2   \right] \hat{z},\nonumber\\
 &=& \mu_B  \left[ 1+ \frac{1}{2} \kappa (z\alpha)   - \frac{1}{3} (z\alpha)^2 \right] \hat{z}
 \label{m_self}  
\end{eqnarray}
where we have gone back to (\ref{betasquare}) and (\ref{Nsquare}), expressed $\beta^2$ and $N^2$ in terms $\alpha$ and $z$, and substituted for $\frac{1}{2}\bar{a_0}\bar{\alpha}e = \mu_B$,  the Bohr magneton. 
The only approximation in  (\ref{m_self}) is in the right hand side of the last equality where we have  made the replacement, 
 $$ \beta^2=(z\alpha)^2+\frac{1}{4}(z\alpha)^4 +{\cal O}(z\alpha)^6 \approx (z\alpha)^2.$$
 Numerically the difference is in the sixth decimal place.
\\
 
\subsection{Determining the coupling constants, $\kappa$ and $z$}
As a refresher  let us first  have a look at the gyromagnetic ratio of the electron as formulated in QED, and measured in the laboratory.
There is a proportionality  between the  spin and/or the orbital angular momentum, $S/\hbar$, and the magnetic moment, $\mu/\mu_B$, of a magnetized particle, see e.g. \citeyear{jackson1999classical}, 
\begin{eqnarray}
\frac{\mu}{\mu_B}= g \frac{S}{\hbar},~~~\mu_B = \frac{e\hbar}{2m_e c}=\frac{1}{2}a_0 \alpha e, \textrm{Bohr magneton},  \label{gfactor}
\end{eqnarray} 
where $g$ is a dimensionless constant known as the g-factor of the particle.  
Laboratory measurements of \cite{hanneke2008new} and  theoretical QED derivations of \cite{aoyama2012complete} of the anomalous magnetic moment of the electrons and muons agree to 12 decimal places and are considered as the  stringiest test of the validity of QED. A comprehensive and uptodate  review of lepton gyromagnetic ratios can be found in the recent arXiv article of \cite{quigg2021notes}. For our reference below, it suffices to quote the following approximate $\alpha$-dependent expression abstracted from \citeyear{aoyama2012complete} and \citeyear{aoyama2015tenth}: 
\begin{eqnarray}
g = 2\left[1+\frac{1}{2}\left(\frac{\alpha}{\pi}\right) - 0.328~ 478 \left(\frac{\alpha}{\pi}\right)^2+ 1.181~241 \left(\frac{\alpha}{\pi}\right)^3
 \cdots\right].  \label{gAoyama}
 \end{eqnarray}
 The $\alpha/2\pi$ term is the celebrated 1-loop calculation of Schwinger (1948).  It was also obtained by the pre- QED quantum physics by adding the $\kappa$  term of (\ref{LDirac}) to the Pauli - Schroedinger or to the Dirac Hamiltonian. 
The term $0.328 (\alpha/\pi)^2\approx 1.913\times10^{-6}$ comes mainly from the 2- loop  QED calculations. There are, however,  small contribution of the order $10^{-6}$ to it from the electroweak and the hadronic light by light interactions.   

  Let us now look at the g-factor associated with the  self induced magnetic moment of (\ref{m_self}). By definition of (\ref{gfactor}) and noting that the electron spin is $\frac{1}{2}$, one gets,
\begin{eqnarray} \frac{\mu_e}{\mu_B}/ \frac{1}{2} = 
g_{e} = 2 \left[ 1+ \frac{1}{2} \kappa z\alpha - \frac{1}{3} z^2\alpha^2  \right]. \label{gse}
\end{eqnarray} 
It only suffices to let
\begin{eqnarray}
\kappa =1 ~~~ \textrm{and} ~~~ z =\frac{1}{\pi}, \label{kappazcouplings}
\end{eqnarray}
and arrive at 
\begin{eqnarray}
g_{e} = 2\left[1+\frac{1}{2}\left(\frac{\alpha}{\pi}\right) - \frac{1}{3} \left(\frac{\alpha}{\pi}\right)^2 \right].  \label{gself}
\end{eqnarray}
 
  The odd power $\frac{1}{2}(\frac{\alpha}{\pi})$  in (\ref{gself}) comes from the $\kappa$ coupling in (\ref{LDirac}) and  is the same as that of  Schwinger but without invoking the QED formalism. It was already known by 1930's that the Pauli interaction was able to account for the $\alpha$-order anomalous magnetic moment of the electron; but it was set aside on the grounds that it was an ad-hoc addition; see e.g. \citeyear{sakurai2006advanced}. Here we argue that the term $\frac{1}{2}(\frac{\alpha}{\pi})$  is a symmetry based and a Noether invariant of the electron coming from the symmetry of  (\ref{LDirac}) under the restricted Lorentz gauge. 
  
  The even power $-\frac{1}{3}(\frac{\alpha}{\pi})^2$ is the contribution from the conventional  minimal coupling in (\ref{LDirac}). It is also a symmetry based and a Noether invariant under the minimal coupling symmetry of (\ref{LDirac}).  Numerically, it differs from  $-0.328(\frac{\alpha}{\pi})^2$   by only $1.5\%$, which can be  easily attributed  to the   electroweak and hadronic interactions in the latter.

 In further cycles of iteration one expects the pattern to repeat itself. That is, in an expansion of the g-factor in powers of $\alpha/\pi$, the $\kappa$ coupling to be responsible for the odd power contributions and the minimal coupling to give the even power ones.

 Considering all these reservations, we think  the choice of (\ref{kappazcouplings}) for the coupling constants is in good  agreement with  (\ref{gAoyama}). We present it as a support to  what we have been conjecturing so far, whose main ingredients are:  a)   the  mutual interaction of the two QW and EM fields associated with a charged particle,  and b) the restricted Lorentz gauge as an essential component of the U(1) gauge symmetry. 
 
  \subsection{Calculation of the self- induced magnetic field} 
Integration of (\ref{nablaAphi}) is  straightforward, but lengthy. Details are given in Appendix A.  Here we only quote the far distance limit of the magnetic field from (\ref{Brinfty}) and (\ref{Bthetainfty}),  
 
   \begin{eqnarray}
       B_r &\rightarrow & \mu_B \left[ 1+ \frac{1}{2}\left(\frac{\alpha}{\pi}\right) 
    - \frac{1}{3} \left(\frac{\alpha}{\pi}\right)^2  \right] \frac{2\cos\theta}{r^3}, ~~\textrm{as}~~ r \rightarrow \infty, \label{Brinftybis} \\
     B_\theta &\rightarrow & \mu_B \left[ 1+ \frac{1}{2}\left(\frac{\alpha}{\pi}\right) 
    - \frac{1}{3} \left(\frac{\alpha}{\pi}\right)^2  \right] \frac{\sin\theta}{r^3}, ~~\textrm{as}~~ r \rightarrow \infty, \label{Bthetainftybis}
   \end{eqnarray}
   The far distance field is a magnetic dipole in which the Bohr  magneton of the electron is modified by a g- factor correct to $(\frac{\alpha}{\pi})^2$ order.  

\section{Concluding remarks}
We have  addressed three  interrelated concepts in quantum and non quantum physics;  i. the assumption of point particles in non- quantum physics; ii. the one-sided action of EM  on quantum wave fields  without invoking a re-action from the latter;  iii.  the left out restricted Lorentz gauge  from the U(1) gauge symmetry in quantum physics.\\

\noindent
To avoid the first two predicaments we have  argued that there are two fields associated with each charged particle, an EM field and a QW field. We have conjectured a mutual action-reaction partnership between the two.  
\begin{itemize}
\item
This provision makes the QW  field to shares its singularity free feature  with the EM field, remove  its Coulomb-like singularities and along with the ensuing divergent integrals, 
\item
Electron's charge gets distributed. A spinning distributed charge, however, is an electric current system. Therefore, the electron acquires a self induced magnetic moment.  
\end{itemize}

As to the third case, we have  proposed to enlarge the U(1) symmetry, by requiring invariance of the EM and QW fields under the restricted Lorentz gauge. This provision elevates the  old anomalous Pauli interaction, into a  symmetry based interaction and invites in the $\kappa$ coupling in addition to the conventional minimal $z$ coupling. \\

The resulting dynamical equations from the Lagrangian of (\ref{LDirac}) for the Dirac and the EM fields are coupled together and are,  therefore, non-linear. In an iteration scheme we have analysed them beginning with the ground state of the  Pauli-Dirac equation as the starting step. Deviations from  the classical results, show up at distances of the order of Compton wavelength, $\approx 10^{-12}$ m from the origin. \\

By Noether's theorems, symmetries and constants of motion are synonymous.  
 In the fist round of iteration, the two constants of motion associated with the $\kappa$ and $z$ symmetries are the  $\frac{1}{2}(\frac{\alpha}{\pi})$  and the $-\frac{1}{3}(\frac{\alpha}{\pi})^2$ contributions   to the anomalous g- factor of the electron, respectively. That the first is exactly the same and the second differs by only $1.5 \% $ from the QED- derived and the laboratory- measured  values is striking. We present them as a support to what we have conjectured in this paper.  \\

In higher steps of iteration of (\ref{psiDirac}) and (\ref{FmunuDiracCoupled}) we anticipate the pattern of contributions to the g- factor to repeat itself; the odd powers of $(\frac{\alpha}{\pi})$  to come from the $\kappa$ coupling and the even powers to result from the $z$ coupling.
\\

  Commonly, a free electron is described as a physical entity with a given charge, mass and spin. If exposed to a magnetic field it is also apt to acquires a spin magnetic moment.  Otherwise, speaking of the  magnetic moment of a free electron is not a common parlance. The picture we are presenting here is somewhat different. 
A free electron with a spinning  distributed charge is  a current system as well. Thereof, it has a distributed magnetic moment density and a total magnetic moment  inseparable from it.   \\

 We have not second- quantized the Dirac field  and have not used the QED formalism. Throughout the text, however, we have talked of commonalities between what we have done here and the QED scenarios.  For examples,  our distributed charge density and the scale invariance of the QED both serve to get rid of  the point charge assumption; our action-reaction conjecture and the re-normalization scenario of QED both are provisions to avoid divergent integrals; our logarithmic singularities in the charge and the magnetic moment densities have counterparts in QED; etc. There must further similarities and deeper connections between the two approaches worth looking into. \\

  Last but not the least. 
In Maxwell's equations, the electric and magnetic fields play dual roles. So much so, that one may devise a duality transformation to combine them and/or replace one by the other  \citep{jackson1999classical}.  Is there such a duality in the mutually coupled EM-QW fields  of our proposed electron? Does any  concept relating to its magnetic property has a counterpart in its electric characteristics and vice versa?   
For instance, is there a counterpart to the anomalous magnetic g- factor to be called  an 'anomalous electric g-factor'? 
Here is a suggestion; in (\ref{Brinfty}) and (\ref{Bthetainfty}), the magnetic dipole field of the  Bohr magneton is modified by the anomalous magnetic g-factor. Can one similarly coin the term and say that the factor $(1- \beta^2/4)$ in (\ref{A0DiracInfty})  is the anomalous electric g-factor that  modifies  the monopole Coulomb field of the electron. The list of parallelisms is long. A comprehensive study of it  should be worth of its while.

 \subsubsection{What we haven't done:}
 
 Mathematical analysis and computations have been taxing. 
We have attempted only the first round of iteration and have obtained the g-factor correct to the order $(\alpha/\pi)^2$. Further rounds of iteration should give higher power  corrections, and  better comparisons with QED and observations. \\

We have analysed only the ground state of the Pauli-Dirac equation (\ref{HDirac}).  Iterations with  higher order eigenstates should be interesting.\\

Iterations beginning with eigenstates  end up with static electric and magnetic fields. To have time dependent EM,  one may attempt linear combinations of the eigenstates.\\

Analysis of the spectrum of hydrogen- like atoms with the  hitherto proposed electron,  with a singular positive charge sitting at the center  should be of interest. \\

All sorts of scattering with the hitherto proposed electron are available  for contemplation. The list may include  electron-molecule scattering, Compton scattering,   deep inelastic scattering, etc. \\

 Second quantization of (\ref{psiDirac}) and (\ref{FmunuDiracCoupled}) and  eventually their QED implications is in our agenda. \\

\noindent 
 \textbf{Acknowledgement:} I am indebted to H. Fazli and M. H. Vahidinia for their constructive comments. I also thank R. G. Meimanat for proofreading the manuscript.

\section{Appendix A. Calculation of the vector potential and the magnetic field}
The integral of the Poisson  equation (\ref{nablaAphi}) is,
\begin{eqnarray}
&&\mathbf{A}(u, \theta)  
    = \frac{N^2\beta^2}{3\pi \bar{a}_0\bar{\alpha}} e   
                   \int \frac{1}{|\mathbf{u}-\mathbf{u'}|}   \label{Aphi} \\ 
  &&  \hspace{.5cm}  \times  \left[(1+\frac{1}{2}\kappa\bar{\alpha}) e^{-u'}u'^{-\beta^2} 
                  + \frac{1}{2}\kappa \bar{\alpha}\beta^2 e^{-u'}u'^{-(!+\beta^2)}   
                          \right]\hat{\phi}' u'^2 du' \sin \theta'   d\Omega'. \nonumber
  \end{eqnarray} 
  In the spherical harmonic expansion of $|\mathbf{u}-\mathbf{u}'|^{-1}$ only the first order harmonics, $\frac{u_<}{u_>^2}Y_1^m(\theta, \phi), m=\pm 1$, contribute to the integral. Further reduction of (\ref{Aphi}) gives,
    \begin{eqnarray}
 && \mathbf{A}(u, \theta) = \frac{N^2\beta^2}{3\pi \bar{a}_0\bar{\alpha}} e  
        \left[  (1+ \frac{1}{2}\kappa\bar{\alpha} ) A^{(1)}(u) 
        + \frac{1}{2} \kappa \bar{\alpha}\beta^2 A^{(2)}(u)\right]\frac{\sin\theta}{u^2}\hat{\phi}
       \nonumber \\ \label{AvecDirac}.
        \end{eqnarray}
  where
  \begin{eqnarray}
A^{(1)}(u) &=&  \Gamma(4-\beta^2, u) -u^3 \Gamma(1-\beta^2, u) + u^3\Gamma(1-\beta^2) 
\label{A1(u)}\\
&\rightarrow &~~~ 6(1- 4\beta^2/3), ~~  u\rightarrow \infty, \nonumber \\
&\rightarrow &  - \frac{3}{4}(1+ 5\beta^2/4)u^{(4-\beta^2)} 
+ u^3(1 +\beta^2/2),   ~~ u\rightarrow 0,  \nonumber\\
\nonumber\\
A^{(2)}(u) &=&  \Gamma(3-\beta^2, u) -u^3 \Gamma(-\beta^2, u) + u^3\Gamma(-\beta^2),  \label{A2(u)} \\
&\rightarrow & 2(1 - \beta^2),  ~ ~ u\rightarrow \infty, \nonumber \\
&\rightarrow &  \frac{1}{\beta^2}(1+ \beta^2/3)u^{(3-\beta^2)}, ~~ u\rightarrow 0, \nonumber
\end{eqnarray}
The resulting self induced magnetic field, 
$$\mathbf{B}= \nabla \times \mathbf{A}=\frac{2}{\bar{a}_0}\nabla_u \times \mathbf{A},$$
 is in the meridional plane,  
 \begin{eqnarray}
&& B_r=~~ \frac{N^2\beta^2}{ 6\pi \bar{a}_0^2\bar{\alpha}} e  
        \left[  (1+ \frac{1}{2}\kappa\bar{\alpha} ) A^{(1)}  
        + \frac{1}{2} \kappa \bar{\alpha}\beta^2 A^{(2)} \right]\frac{2\cos\theta}{u^3}
       \label{BrDirac} \\
&& B_\theta = - \frac{N^2\beta^2}{ 6\pi \bar{a}_0^2\bar{\alpha}} e     \label{BthetaDirac}\\
&& ~~~~\times
       \left[(1+ \frac{1}{2}\kappa\bar{\alpha} ) 
     ~ u^2\frac{\partial}{\partial u}\left(\frac{A^{(1)}}{u}\right)  
     + \frac{1}{2} \kappa \bar{\alpha}\beta^2 
       ~u^2\frac{\partial}{\partial u}\left(\frac{A^{(2)}}{u}\right) 
         \right]  \frac{\sin\theta}{u^3}.\nonumber
    \end{eqnarray}
   The angular dependence of the self induced magnetic field, $2\cos\theta$ and $\sin\theta$ in (\ref{BrDirac}) and (\ref{BthetaDirac}), is that of a dipole field, but its $r$- ($u$-) dependence is not. Their far distance limit, however, is a genuine dipole with the anomalously modified g-factor of  (\ref{gself}), 
   \begin{eqnarray}
       B_r &\rightarrow & \mu_B \left[ 1+ \frac{1}{2}\left(\frac{\alpha}{\pi}\right) 
    - \frac{1}{3} \left(\frac{\alpha}{\pi}\right)^2  \right] \frac{2\cos\theta}{r^3}, ~~\textrm{as}~~ r \rightarrow \infty, \label{Brinfty} \\
     B_\theta &\rightarrow & \mu_B \left[ 1+ \frac{1}{2}\left(\frac{\alpha}{\pi}\right) 
    - \frac{1}{3} \left(\frac{\alpha}{\pi}\right)^2  \right] \frac{\sin\theta}{r^3}, ~~\textrm{as}~~ r \rightarrow \infty, \label{Bthetainfty}
   \end{eqnarray}
   We again emphasize that, thanks to the mutual interaction between EM and the quantum wave fields,  there are no singularities in the EM potentials and fields nor any divergent integral relating to them. Logarithmic divergences encountered before are in the charge and current densities, again reminiscent of the scaling symmetry practised in QED.\\

\section{Appendix B.   $\Gamma$ functions}
\subsection{Complete $\Gamma$ functions}

A complete $\Gamma$ function is defined as
\begin{eqnarray}
\Gamma(s+1) &=&  s\Gamma(s) =  \int_0^\infty \exp(-u) u^s du, \label{Gamma.comp}
\end{eqnarray}
 where, in this paper $s$ is a    real number. The first equality is the recursion relation; the second is the definition. For an integer $n$ one has,
\begin{eqnarray}
\Gamma(n+1) &=& n! \nonumber
\end{eqnarray}
For $n-\beta^2, ~~\beta^2\ll 1$, one may  expand complete $\Gamma$'s in $\beta^2$. 
 Up to order $\beta^2$ one finds,
 \begin{eqnarray}
\Gamma(1-\beta^2) &=& 1+ \beta^2/2, \nonumber\\  
\Gamma(2-\beta^2) &=& 1-\beta^2/2, \nonumber\\ 
\Gamma(3-\beta^2) &=& 2(1-  \beta^2), \nonumber\\ 
\Gamma(4-\beta^2) &=& 6(1-4\beta^2/3). \label{GamCompEx}
\end{eqnarray}
 The expressions above are calculated  by Wolfram Mathematica. They can also be read from the asymptotes of  Figure \ref{fig.GamDerivs}. The recursion relation gives,
 \begin{eqnarray}
 \Gamma(n+1-\beta^2) &=& (n-\beta^2)\Gamma(n-\beta^2). \label{recur.Gam.alp2}
 \end{eqnarray}
  
\subsection{Incomplete $\Gamma$ functions} 

\begin{figure}[H]
\centering
\includegraphics[scale=.5]{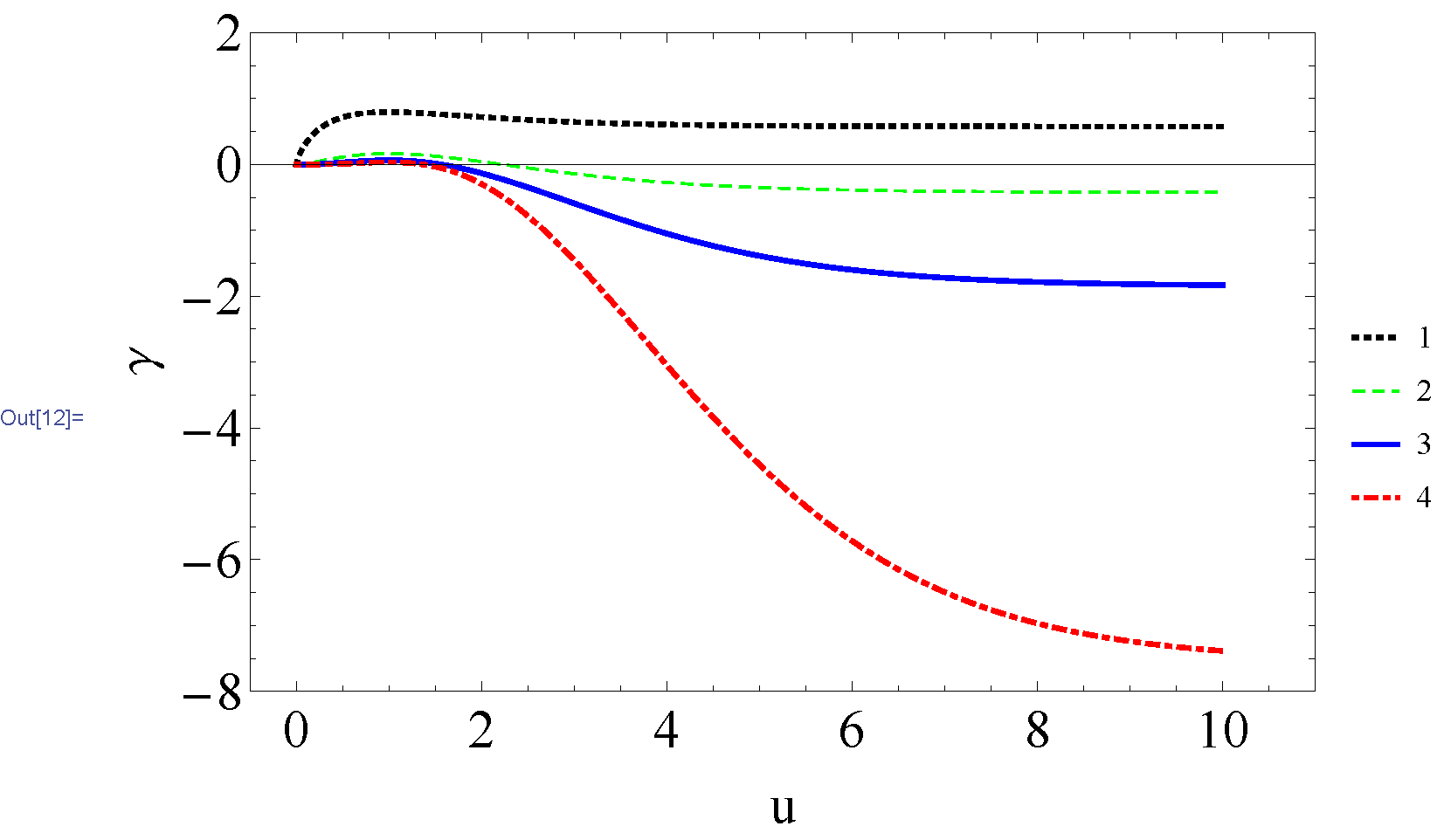}
 \caption{Plots of $\gamma(n+1, u)$'s as functions of u. They tend to $0$ as $u$ tends to $0$. and tend to flat asymptotes as $u$ tends to $\infty$.}
 \label{fig.GamDerivs}
\end{figure}

An incomplete $\Gamma$   is defined as
\begin{eqnarray}
\Gamma(s+1,u)&=& s\Gamma(s,u) - \exp(-u) u^s = \int_0^u \exp(-u) u^s du.  \label{Gamma.s.u}
  \end{eqnarray}
   The first equality is the recursion relation; the second is the definition.
If $n$ integer,  incomplete $\Gamma$'s  have analytical solutions. Examples are,
\begin{eqnarray}
\Gamma(1, u) &=& 1-e^{-u}\\
\Gamma(2, u) &=& 1-\left(1+u\right) e^{-u}\\
\Gamma(3, u) &=& 2\left[1-\left(1+u +\frac{1}{2} u^2\right) e^{-u}\right]\\
\Gamma(4, u) &=& 6\left[ 1-\left(1+u +\frac{1}{2} u^2+\frac{1}{6} u^3\right) e^{-u}\right]  
 \end{eqnarray}
Limiting behaviours of incomplete $\Gamma$'s as $u\rightarrow 0 ~\textrm{ or }\infty$ are as follows,
\begin{eqnarray}
\Gamma(s+1,u)&\rightarrow &  u^{s+1}/(s+1)~\textrm{as}~u\rightarrow 0, \\
                       &\rightarrow &\Gamma(s+1)~~~~~~~~\textrm{as}~u\rightarrow \infty. 
\end{eqnarray}
For $n-\beta^2, ~~\beta^2 \ll 1$, one may  expand incomplete $\Gamma$'s in $\beta^2$. Writing 
$$u^{-\beta^2}= \exp(- \beta^2\ln u) = 1 - \beta^2\ln u ,$$
up to order $\beta^2$ one finds,
 \begin{eqnarray}
\Gamma(n+1-\beta^2,u) &=& \Gamma(n+1,u) +\beta^2 \gamma(n+1,u), \label{GammaTaylorExp}
\end{eqnarray}
where
\begin{eqnarray}
  \gamma(n+1, u) &=&  -\int_0^u e^{-u} u^n \ln u du, \label{gamma}\\
        &\rightarrow & -\frac{u^{n+1}}{(n+1)}\left[\ln u-\frac{1}{(n+1)}\right]~
         \textrm{as}~ u\rightarrow 0, 
        \nonumber\\
        &\rightarrow & \textrm{flat asymptotes as } u\rightarrow \infty.\nonumber
 \end{eqnarray}
Examples of $\gamma(1,u\rightarrow\infty)$ are,
 \begin{eqnarray}
   \gamma(1,\infty)= -1, ~~\gamma(2,\infty) = 1,~~\gamma(3,\infty) =  2,~~\gamma(4,\infty) = 8.\label{GamDerivAsymp}
\end{eqnarray}
  For $n=1,2,3,4$, $\gamma(n+1,u)$'s are plotted in Figure\ref{fig.GamDerivs}
for $u\rightarrow 0$. Direct integration of $\Gamma(n+1-\beta^2, u )$ gives,
\begin{eqnarray}
\Gamma(4-\beta^2) &=& \frac{1}{4-\beta^2}u^{(4-\beta^2)}, ~~\textrm{as}~~u\rightarrow 0, \label{4betau}\\
\Gamma(3-\beta^2) &=& \frac{1}{3-\beta^2}u^{(3-\beta^2)}, ~~u\rightarrow 0, \label{3betau}\\
\Gamma(2-\beta^2) &=&\frac{1}{2-\beta^2}u^{(2-\beta^2)},~~u\rightarrow 0,\label{2betau} \\
\Gamma(1-\beta^2) &=& \frac{1}{1-\beta^2}u^{(1-\beta^2)}, ~~~u\rightarrow 0. \label{1betau}
\end{eqnarray}


%

\bibliographystyle{apalike} 
\bibliography{BibliographyFile}

\begin{thebibliography}{}

\bibitem[Aoyama et~al., 2012]{aoyama2012complete}
Aoyama, T., Hayakawa, M., Kinoshita, T., and Nio, M. (2012).
\newblock Complete tenth-order qed contribution to the muon g- 2.
\newblock {\em Physical review letters}, 109(11):111808.

\bibitem[Aoyama et~al., 2015]{aoyama2015tenth}
Aoyama, T., Hayakawa, M., Kinoshita, T., and Nio, M. (2015).
\newblock Tenth-order electron anomalous magnetic moment: Contribution of
  diagrams without closed lepton loops.
\newblock {\em Physical Review D}, 91(3):033006.

\bibitem[Blinder, 2003]{blinder2003singularity}
Blinder, S. (2003).
\newblock Singularity-free electrodynamics for point charges and dipoles: a
  classical model for electron self-energy and spin.
\newblock {\em European journal of physics}, 24(3):271.

\bibitem[Born and Infeld, 1934]{born1934foundations}
Born, M. and Infeld, L. (1934).
\newblock Foundations of the new field theory.
\newblock {\em Proceedings of the Royal Society of London. Series A, Containing
  Papers of a Mathematical and Physical Character}, 144(852):425--451.

\bibitem[Einstein, 2014]{einstein2014meaning}
Einstein, A. (2014).
\newblock {\em The meaning of relativity: Including the relativistic theory of
  the non-symmetric field}, volume~32.
\newblock Princeton University Press.

\bibitem[Green, 1947]{green1947self}
Green, A.~E. (1947).
\newblock Self-energy and interaction energy in podolsky's generalized
  electrodynamics.
\newblock {\em Physical Review}, 72(7):628.

\bibitem[Hanneke et~al., 2008]{hanneke2008new}
Hanneke, D., Fogwell, S., and Gabrielse, G. (2008).
\newblock New measurement of the electron magnetic moment and the fine
  structure constant.
\newblock {\em Physical Review Letters}, 100(12):120801.

\bibitem[Jackson, 1999]{jackson1999classical}
Jackson, J.~D. (1999).
\newblock Classical electrodynamics.

\bibitem[Land{\'e} and Thomas, 1941]{lande1941finite}
Land{\'e}, A. and Thomas, L.~H. (1941).
\newblock Finite self-energies in radiation theory. part ii.
\newblock {\em Physical Review}, 60(7):514.

\bibitem[Lazar and Leck, 2020]{lazar2020second}
Lazar, M. and Leck, J. (2020).
\newblock Second gradient electrodynamics: A non-singular relativistic field
  theory.
\newblock {\em Annals of Physics}, 423:168330.

\bibitem[Podolsky, 1948]{podolsky1948228}
Podolsky, B. (1948).
\newblock 228. b. podolsky and p. schwed.
\newblock {\em Rev. Mod. Phys}, 20:40.

\bibitem[Quigg, 2021]{quigg2021notes}
Quigg, C. (2021).
\newblock Notes on lepton gyromagnetic ratios.
\newblock {\em arXiv preprint arXiv:2105.07866}.

\bibitem[Sakurai, 2006]{sakurai2006advanced}
Sakurai, J.~J. (2006).
\newblock {\em Advanced quantum mechanics}.
\newblock Pearson Education India.

\bibitem[Santos, 2011]{santos2011plasma}
Santos, R. B.~B. (2011).
\newblock Plasma-like vacuum in podolsky regularized classical electrodynamics.
\newblock {\em Modern Physics Letters A}, 26(25):1909--1915.

\bibitem[Schiff, 1955]{schiff1955quantum}
Schiff, L. (1955).
\newblock Quantum mechanics mcgraw-hill.
\newblock {\em New York}, 19752:267--280.

\end{thebibliography}

\end{document}